\documentclass[useAMS,usenatbib]{mn2e}
\usepackage{graphicx}
\title[4U 1636--53: I. Long-term evolution and kHz QPOs]
{RossiXTE monitoring of 4U 1636--53: I. Long-term evolution and kHz Quasi-Periodic  Oscillations}
\author[T. Belloni et al.]{T. Belloni$^{1}$
\thanks{E-mail:tomaso.belloni@brera.inaf.it}, 
J. Homan$^{2}$,
S. Motta$^{1,3}$, E. Ratti$^{1,3}$,
M. M\'endez$^{4,5,6}$\\
$^{1}$INAF-Osservatorio Astronomico di Brera, Via E. Bianchi 46, I-23807 Merate
(LC), Italy\\
$^{2}$Center for Space Research, Massachusetts Institute of Technology, 77 Massachusetts Avenue, Cambridge, MA 02139-4307, USA\\
$^{3}$Universit\`a di Milano Bicocca, Piazza della Scienza 3, 20126 Milano, Italy\\
$^{4}$SRON, Netherlands Institute for Space Research, Sorbonnelaan 2, 3584 CA Utrecht, The Netherlands\\
$^{5}$Astronomical Institute `Anton Pannekoek', University of Amsterdam,
Kruislaan 403, 1098 SJ, Amsterdam, the Netherlands\\
$^{6}$Astronomical Institute, University of Utrecht, PO Box 80000, 3508 TA
Utrecht, The Netherlands\\
}

\begin{document}

\date{
Accepted 2007 May 3. Received 2007 May 2}

\pagerange{\pageref{firstpage}--\pageref{lastpage}} \pubyear{2006}

\maketitle

\label{firstpage}

\begin{abstract} We have monitored the atoll-type neutron star
low-mass X-ray binary 4U 1636--53 with the Rossi X-Ray Timing Explorer (RXTE)
for more than 1.5 years. Our campaign consisted of short
($\sim$2 ks) pointings separated by two days, regularly monitoring the spectral
and timing properties of the source.
During the campaign we observed a
clear long-term oscillation with a period of $\sim$30-40 days,
already seen in the light curves from the RXTE All-Sky Monitor, which
corresponded to regular transitions between the hard (island) and soft
(banana) states. We detected kHz QPOs in about a third of the
observations, most of which were in the soft (banana) state. The
distribution of the frequencies of the peak identified as the lower
kHz QPO is found to be different from that previously
observed in an independent data set. This
suggests that the kHz QPOs in the system shows no intrinsically
preferred frequency.

\end{abstract}


\section{Introduction}

High-frequency quasi-periodic oscillations (QPO) in neutron-star
X-ray binaries often appear in pairs, with frequencies ranging from a
few hundred Hz to more than 1 kHz (see van der Klis 2006 for a
review). These so-called kHz QPOs, which were discovered with the Rossi
X-Ray Timing explorer (RXTE), provide a probe into the accretion flow
very close to the compact objects and can possibly serve as a tool to
observe effects of general relativity. Their frequencies are strongly
correlated with other timing and spectral features (see Ford \& van
der Klis 1998; Psaltis, Belloni \& van der Klis 1999; Belloni,
Psaltis \& van der Klis 2002; van Straaten et al. 2005). In
particular, the frequencies of the lower ($\nu_{low}$) and upper
($\nu_{high}$) kHz QPO are strongly correlated with each other (see
Belloni, M\'endez \&  Homan 2005, hereafter BMH05). The correlation
is roughly linear (but see Zhang et al. 2006) and similar for all
sources (Belloni, M\'endez \& Homan 2007).

A number of theoretical models have been proposed for the interpretation of these oscillations, based on the identification of their frequencies with various characteristic frequencies in the inner accretion flow (Stella \& Vietri 1999; Osherovich \& Titarchuk 1999; Zhang 2004; Lamb \& Miller 2003). However, there is still no consensus as to the origin of the QPOs. 

In an attempt to determine the possible presence of
preferred frequencies, as predicted by certain resonance models
(see e.g. Abramowicz et al. 2003), BMH05 investigated the distribution of the kHz QPO frequencies
for separate sources. They analyzed large samples of RXTE data of
five systems and found indications for preferred
frequencies in each of the sources. However, it is known that the
time evolution of kHz QPO frequencies is similar to a random walk,
i.e. the current frequency cannot jump arbitrarily away from the one
from a few dozen seconds before (see van der Klis 2006). BMH05 showed
that a random walk in frequency, if not sampled for a long time
compared to the random walk time scale, also produces apparent preferred
frequencies. 

%-----------------------------------------------------------------------
\begin{figure*}
\begin{center}
\includegraphics[angle=0,width=18cm]{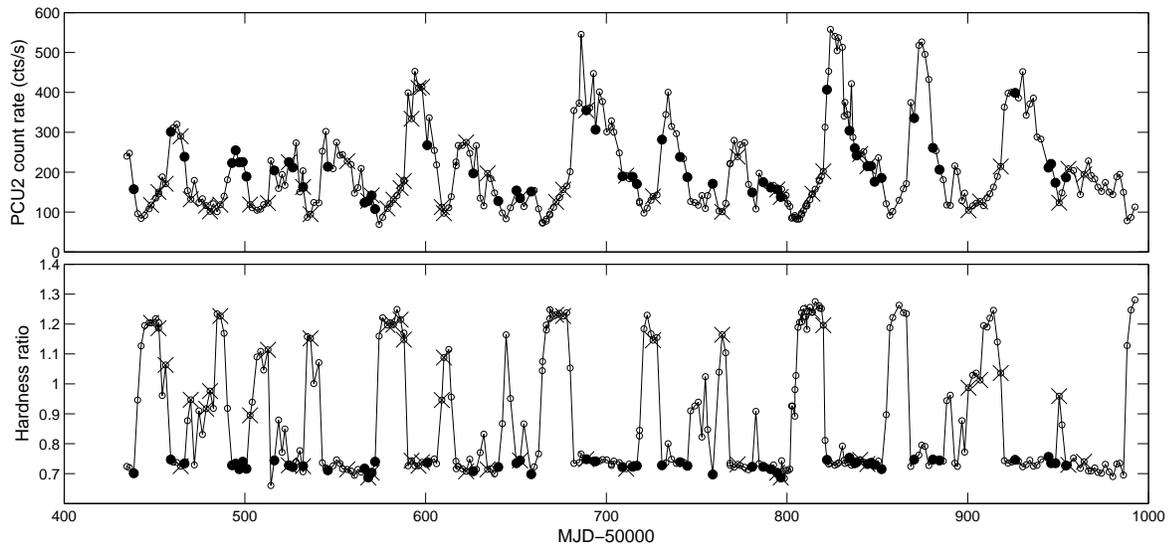}
\end{center}
\caption{Top panel: long-term PCA light curve of 4U 1636--53
(one point per observation) in the 2-60 keV energy band. Filled circles correspond to observations where we detected a narrow high-frequency QPO which we identified as a lower kHz QPO. Crosses indicate the detection of X-ray bursts.
Bottom panel: corresponding evolution of the hardness ratio. The meaning of the symbols is the same as in the top panel.}
\label{licu}
\end{figure*}
%-----------------------------------------------------------------------

To obtain a data set with an improved sampling of the kHz QPO
frequency evolution and to determine an unbiased frequency
distribution  we started a monitoring campaign of one of the five
sources studied by BMH05,  4U 1636--53. This source is a member of
the atoll class (see Hasinger \& van der Klis 1989). It contains a
neutron star accreting matter from a companion star of mass $\sim
0.4$ $M_\odot$   with an orbital period of 3.8 hr (see, e.g., van
Paradijs et al.\ 1990). Type-I X-ray bursts were observed several
times from  4U 1636--53, sometimes with unusual shapes (e.g.\ Turner
\& Breedon 1984; van Paradijs et al.\ 1986), or unusually long
durations (e.g.\ Wijnands 2001). New results on the X-ray bursts have
recently been presented  by Bhattacharyya \& Strohmayer (2005,2006).
A long-term quasi-periodic variability has been
reported from All-Sky Monitor (ASM) data (Shih et al. 2005).  

Two simultaneous kHz QPOs have been observed in 4U 1636--53, with
rms   amplitudes increasing with photon energy (Zhang et al.\ 1996;
Wijnands et al.\ 1997).  The frequency of the lower kHz QPO
increases, and its  amplitude decreases, with inferred $\dot M$ (Di
Salvo, M\'endez \& van der Klis 2003; their data are the same
as those used by BMH05), and at high $\dot M$ the
frequency difference between the two QPOs is in the range $240-280$
Hz, significantly lower than half the frequency of the oscillations
detected during type-I bursts (M\'endez et al.\ 1998).  It has been
shown that at low inferred mass accretion rate the kHz QPO peak
separation in 4U 1636--53 is $\sim 323$ Hz, which exceeds half  the
neutron star spin frequency as inferred from
burst oscillations (Jonker et al.\ 2002). Third, and possibly fourth,
weaker kHz QPOs have been discovered  simultaneously with the
previously known kHz QPO pair (Jonker et al.\  2000); the new kHz
QPOs are at frequencies of $\sim 58$ Hz above and below
the frequency of the lower kHz QPO respectively, suggesting that
these are sidebands to the lower kHz QPO.   
Barret et al. (2005) reported the results of a systematic archival study of 
the  Proportional Counter Array (PCA)
observations of 4U 1636--53 concentrating on the drop in coherence of
the lower kHz QPO at high frequencies, as previously found by Di Salvo et al.
(2003).  After extension of their analysis to a large sample of sources, 
Barret et al. (2006) interpreted this as evidence for the presence of the
innermost stable orbit around the neutron star. This
conclusion is under discussion (see M\'endez 2006 and Barret et al.
2007).

4U 1636--53, together with 4U~1608--52 and Aql~X-1, also shows QPOs
at mHz  frequencies (Revnivtsev et al.\ 2001).  This feature seems to
occur only in a rather narrow range of mass accretion rates,
corresponding to X-ray  luminosities of $\sim (0.5-1.5) \times
10^{37}$ ergs/s, and they disappear after X-ray bursts. Contrary to
the general behavior of QPOs, the rms amplitude of the mHz QPOs
strongly decreases with energy. These oscillations are thought to be
related to  nuclear burning on the surface of the neutron star (see
Revnivtsev et al.\ 2001).

In this paper we present the results of the first 1.5 years of our
ongoing monitoring campaign of 4U 1636--53. This paper is  devoted to
the frequency distribution of the kHz QPOs, while following articles
will present the results on the X-ray bursts and on a more detailed 
analysis of the overall evolution.

\section{The campaign} 

In March 2005,we started to observe 4U
1636--53 with RXTE for about 2 ks every  two days. Here we include
all observations from 1 March 2005 to 14 September 2006, a total of
305 pointings. During the period 2006 March 3 through April 10, daily
observations were performed. For each dataset, we extracted a
background-subtracted average count rate in the full PCA channel band
(corresponding to $\sim$2--60 keV) using standard FTOOLS. We only
used data form Proportional Counter Unit \#2 (PCU2) 
as it was the only one consistently operational
during all observations. The resulting long-term light curve is shown
in the top panel of Fig. \ref{licu}. An oscillation with a period of 30-40 days (see
Shih et al. 2005) is evident.
A total of 49 X-ray bursts was
detected (marked with crosses in Fig. \ref{licu}). On two occasions,
2005 April 20 and 2005 July 31, two X-ray bursts within the same
observation were detected. A study of the X-ray bursts will be
presented in a forthcoming paper.

\section{Long-term evolution}

\begin{figure}
\begin{center}
\includegraphics[angle=0,width=9cm]{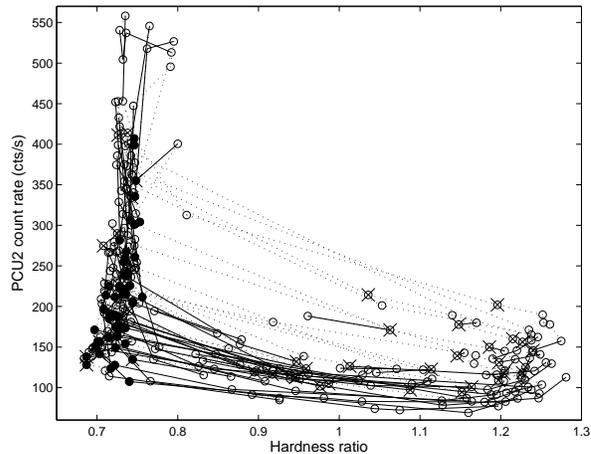}
\end{center}
\caption{Hardness-Intensity diagram of 4U 1636--53 for our campaign, one point per observation. 
Filled circles correspond to observations where we detected a narrow high-frequency QPO which we identified as a lower kHz QPO. Crosses indicate the detection of X-ray bursts.
The lines connect the observations in time sequence: solid line means time evolution from soft
to hard, dotted line the reverse.}
\label{hid}
\end{figure}

In order to follow the X-ray color evolution of the source
along the count rate oscillation (see top panel of Fig. \ref{licu}, we
extracted background-subtracted count rates from the {\it Standard2}
channel bands A=3--10 and B=20--40 (numbering from 1 to 129),
nominally 2.5--5.7 keV and 7.8--16.4 keV respectively.  We then
constructed a hardness ratio $HR = B/A$. 
The time evolution of the hardness ratio can be seen in the bottom 
panel of Fig. \ref{licu}. The count-rate oscillations correspond to strong
spectral changes, which appear to be out of phase with changes in hardness.
In order to study this behaviour in more detail,
we constructed the Hardness-Intensity Diagram (HID),
shown in Fig. \ref{hid}, using the count rate and hardness ratio from Fig. \ref{licu}.
The source
followed a very repetitive path in this diagram, corresponding to the
long-period oscillation visible in Fig. \ref{licu}. Two main regions
can be identified in the HID: the hardest points are all at low count
rate and correspond to the island state of atoll sources. The soft
points span a rather large range of count rate while keeping a nearly
constant hardness: they correspond to the banana branch. In between,
transitional points are observed. In Fig.  \ref{hid} we connected
consecutive observations with a line: a dotted line if the source
moved  from hard to soft (leftward) and a solid line if the source
moved from soft to hard (rightward). It is clear that 4U 1636--53
traveled the diagram describing a counter-clockwise rotation
(although the vertical soft branch in the HID is not actually
followed monotonically from top to  bottom). Also, the density of
transitional points indicate that the transition from hard to soft is
faster than the reverse transition.  The 47 observations with X-ray
bursts are marked with crosses in Fig. \ref{hid}. 
Bursts are observed all over the diagram, although there are none
corresponding to the hardest points of the island state (HR $>$
1.25) and around HR=0.8.

%-----------------------------------------------------------------------
%-----------------------------------------------------------------------

\section{Timing analysis}

For each observation, we divided the light curve from the full PCA
energy band  in intervals 16 seconds long and  produced a Power
Density Spectrum (PDS) for each stretch of data. The time resolution
used was 256$\mu$s, corresponding to a Nyquist frequency of 2048 Hz.
We normalized the PDS according to Leahy et al. (1983), obtaining a
time-frequency image.  We also averaged all PDS corresponding to a
single observation in order to increase statistics. The average PDS
were then normalized to fractional rms (Belloni \& Hasinger 1990).
The frequency region 300--1300 Hz was rebinned into 128 bins 
(frequency resolution $\sim$8 Hz) and the
resulting power spectrum was fitted using {\tt XSPEC 11.3} with a
model consisting of a constant and one or two Lorentzians. In a few
cases, an additional zero-centered Lorentzian  was added to take into
account a broad component. 

In this way we detected a significant QPO peak in 95 observations
from our sample  of 305. In three cases also a second significant
QPO was detected. Since kHz QPO frequencies drift on short time
scales, we rebinned each time-frequency image by a factor of 4 in
time (resolution 64s) and inspected all images for peaks showing
changes in their centroid frequency.  We found such a
variable-frequency peak in 53 cases out of 95, while for the other 42
observations, no significant peak was detected at 64 s resolution.  For the
observations with a visible variable peak, we fitted each 64s PDS
separately with a constant plus Lorentzian model and recovered the
frequency evolution versus time. The fits were performed with a
Lorentzian with a FWHM fixed at 4 Hz in order to constrain the
automatic fits to narrow components. In the following, we refer to
these  as variable-peak observations.

\begin{figure}
\begin{center}
\includegraphics[angle=0,width=9cm]{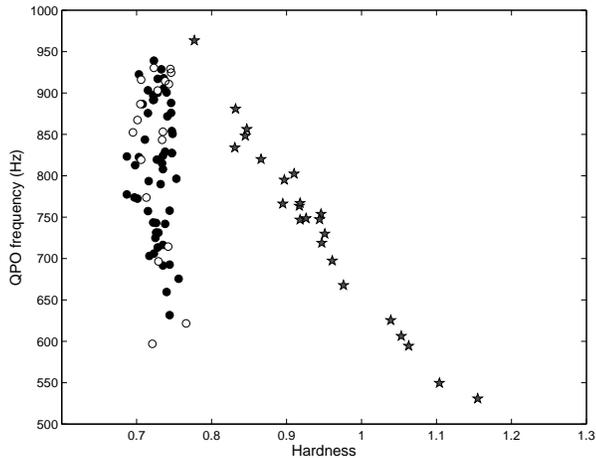}
\end{center}
\caption{Average centroid frequency of the QPO peaks of 4U 1636--53 as a function of spectral 
hardness. Filled circles mark the variable-peak observations,
stars the ones for which the peak is identified as upper kHz QPO. Empty circles correspond to 
non-variable-peak lower-kHz QPOs.
}
\label{q-hardness}
\end{figure}

In the three cases where two peaks were detected, we obviously have a
lower and an upper kHz QPO. When only one peak is detected, the
identification of the QPO has to be made in a  different way. In
Fig.\ \ref{q-hardness}, for each observation we plot the average QPO centroid frequency
as a  function of spectral hardness. Two populations of points
immediately emerge: above a hardness of 0.75, the QPO frequency is
anti-correlated with hardness, while below that threshold there is no
correlation. Such a segregation is known (see M\'endez \& van der
Klis 1999, M\'endez et al. 1999)  and allows us to identify the two
types of QPO: lower-kHz QPO at low hardness, upper-kHz QPO at high
hardness.

%-----------------------------------------------------------------------
%-----------------------------------------------------------------------

\begin{figure}
\begin{center}
\includegraphics[angle=0,width=8cm]{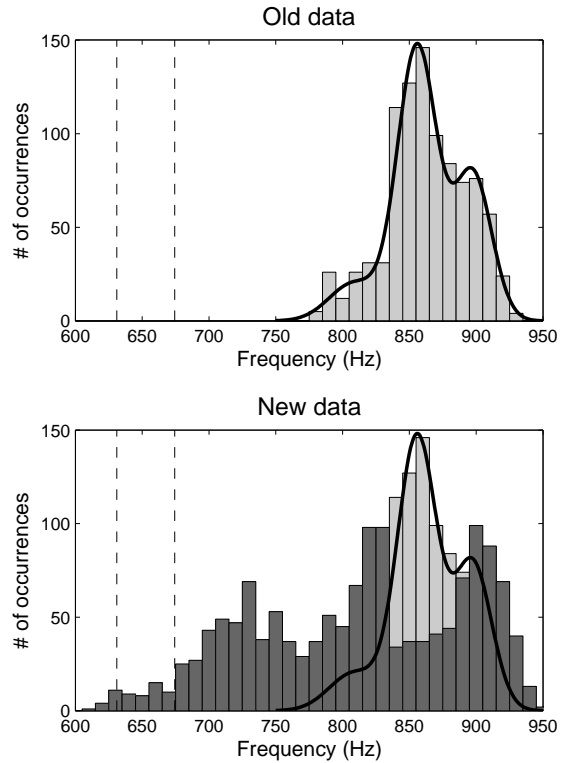}
\end{center}
\caption{Top panel: distribution of frequencies of the lower kHz QPO of 4U 1636--53 
in BMH05. The line is a multi-Gaussian fit to the data (see BMH05).
Bottom panel: same distribution for the variable-peak lower kHz QPO frequencies in this 
work, superimposed to the BMH05 data. The dashed vertical lines indicate the frequency range
that would correspond to a frequency ratio of 3:2 using the correlations of BMH05.
}
\label{distribution}
\end{figure}

With this classification of peaks, we can examine the distribution of
centroid frequencies. The bottom panel of Fig.\ \ref{distribution}
shows the distribution of variable-peak frequencies, together with the
corresponding histogram from BMH05. The new data are distributed in a
very different way and over a wider range of frequencies. There are
broad peaks visible, but none corresponding to the main peak observed
by BMH05. A Kolmogorov-Smirnov test, applied over the same frequency range
gives a probability of 10$^{-26}$ that the
two sets were drawn from the same parent population. The dashed lines
mark the region corresponding, within errors, to frequencies that
would correspond to a 3:2 ratio of upper and lower kHz QPO frequency,
calculated with the frequency correlations in BMH05. While no BMH05
detections fell in that region, a few do so in the new sample. 

\begin{figure}
\begin{center}
\includegraphics[angle=0,width=8cm]{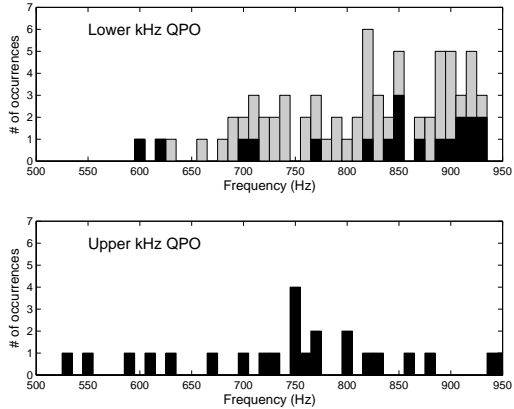}
\end{center}
\caption{Top panel: distribution of the frequencies identified as lower kHz QPO, one average value per observation. Marked in  black are those that could not
be followed at 64s time resolution. 
Bottom panel: distribution of the frequencies identified as upper kHz QPO, one frequency per
observation.
}
\label{dist_broad}
\end{figure}

While the data in Fig. \ref{distribution} represent only the variable-peak
frequencies, i.e. lower kHz QPOs which were sufficiently narrow and
strong to be followed in time with a 64s resolution, it is
interesting to examine the histograms of the remaining frequencies
identified with a lower kHz QPO, consisting of one frequency per
observation, and of the frequencies identified with the upper kHz QPO.
These can be seen in Fig. \ref{dist_broad}. In the top panel are all frequencies
identified with as lower-kHz QPO, one point per observation: in gray the ones which 
could also be followed on 64s time scale, in black the others. In the bottom
panel are all frequencies
identified with as upper-kHz QPO, also one point per observation.
The upper kHz frequencies
are evenly distributed over the full range, while the lower kHz QPOs show
a concentration at high frequencies. 

%-----------------------------------------------------------------------
%-----------------------------------------------------------------------

The distributions described above derive from a set of positive
detections, while a large number of observations did not yield a
significant QPO. This raises a problem of completeness that needs to
be addressed.  We ran simulations to determine the percentage of
lower kHz QPO peaks that we could have missed as a function of
centroid frequency. For each frequency in our range, we produced 100
PDS corresponding to an observation of 2ks with 3 PCUs on and a
source count rate equal to  the minimum observed value (see Fig.
\ref{licu}). The rms and $Q$ value for the QPO were derived from the
curves reported by Barret et al. (2005).  The result is that we are
near 100\% detection efficiency across the frequency range 400-1000
Hz, with a decrease at the boundaries, reaching 30\% at 250 Hz and
1150 Hz. This means that our search is complete within the observed range
and the frequency distributions are not biased for missing additional peaks.

%-----------------------------------------------------------------------
%-----------------------------------------------------------------------

%-----------------------------------------------------------------------
\begin{figure}
\begin{center}
\includegraphics[angle=0,width=9cm]{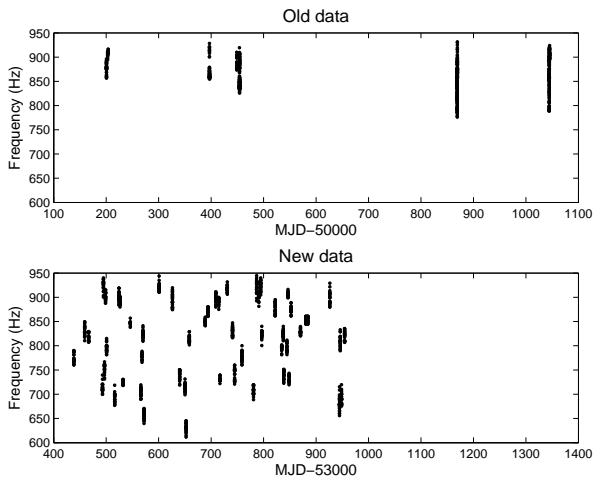}
\end{center}
\caption{Top panel: long term time evolution of the lower kHz QPO frequencies of 4U 1636--53 used by BMH05.
Bottom panel: evolution of the variable-peak lower kHz QPO frequencies from this work. The time
span of the two panels is the same.}
\label{time_history}
\end{figure}
%-----------------------------------------------------------------------

%===========================================================================
\section{Discussion}

Using a series of short observations at regular intervals with RXTE,
we found that the  30 -- 40 day quasi-periodicity seen in the RXTE/ASM
light curve of 4U 1636--53 (Shih et al. 2005) is due to regular state
transitions between the hard (island) and soft (banana) states. These
cyclic state transitions in 4U 1636--53 have been repeting for at least
5 years now, and they follow a hysteresis cycle similar to the one seen
in black-hole binaries in outburst. This is the first time ever that
this hysteresis
pattern has been observed in a persistent neutron star system.

We found that the distribution of frequencies of the lower kHz QPO in
4U 1636--53 changed significantly compared to the distribution in the
observations used by BMH05. This is consistent with the proposal by
BMH05 that the peaks in the distribution of the kHz QPO frequencies
are due to the combination of a random walk of the QPO frequencies and
the observational sampling. This definitely shows that in 4U 1636--53 there are no
preferred frequencies (or frequency ratios) in the distribution of kHz
QPO frequencies.

\subsection{The distribution of QPO frequencies}

In Fig. \ref{distribution}, we compare the distribution of lower kHz QPO frequencies
in 4U 1636--53 from this campaign to the distribution of lower kHz
QPO frequencies for the same source taken from BMH05, which was based on
data taken before 1998. This comparison shows that in the new data the
peaks in the distribution are less pronounced than they were
in the data of BMH05, and are not located at the same frequencies. This
shows that, at least for this source, peaks in the distribution of QPO
frequencies have no special significance but, as suggested by BMH05,
they are due to a combination of the random walk of the QPO frequencies
and the observational sampling.

However, there could be other reasons for the difference in frequency distributions.
         As mentioned above, the simplest interpretation is in the framework of a
	random walk. Sampling a random walk over a limited time span yields
	peaks in the distribution of frequencies even though there are no
	preferred frequencies. This has already been shown by BMH05.

          Even assuming that the frequency variations follow a simple random walk,
	although we know that in
         reality this cannot be strictly a random walk, we need to consider the
	fact that the sampling strategies of the earlier BMH05 data and our new
	data were very different. 
	In Fig. \ref{time_history}, we show the long term frequency evolution 
	(one point every 16 seconds) of the data used in BMH05  compared to
          those of this campaign, using the same time scale on the X axis. The
	BMH05 data consisted of a few long pointings separated by long
	intervals in time, while our regular coverage samples the evolution
	much  better.
The kHz QPOs in neutron-star LMXBs, both Z and atoll sources, are
known to show large ($\Delta\nu > 50$ Hz) frequency changes on
time scales longer than our typical 1/2 hour exposure.  Therefore,
long observations or clusters of observations such as those
implemented in the past are not optimal to obtain an unbiased view
of the QPO evolution. 
Our new campaign yields a  set of observations 
sufficiently spaced so that any correlation between consecutive 
measurements introduced by the random walk of the frequency
is less important.

Finally, there is another possible concurrent cause for the difference
in distribution of frequencies between the two datasets. The data used by
BMH05 correspond to observations in the period 1996--1998, when the 
RXTE/ASM data indicate that 4U 1636--53 was at a rather stable level of
$\sim$20 ASM counts s$^{-1}$ (Shih et al. 2005). After 2000, the overall rate decreased
and the long-term oscillation started (Shih et al. 2005). This means that the
BMH05 frequencies are associated to a high stable flux, while the data from 
our campaign are associated to a much more variable flux level.
It is known that the frequency of the kHz QPOs in these systems show a
positive correlation with source count rate on short (hours) time scales. On long
time scales, this correlation shows shifts in count rate, producing what is
known as ``parallel lines" (M\'endez et al. 1999). A possible explanation for this
phenomenon was proposed by van der Klis (2001): the frequency would depend on 
the current mass accretion rate normalized by its long-term average. 
This would mean that in the BMH05 data, since the long-term average did not vary much,
the frequencies would be observed to vary less than in our more recent data.

The distribution of lower kHz QPO frequencies shown in Fig.
\ref{distribution} gives a different view than what was found by
BMH05, but confirms their conclusions. 
Our new results confirm the
absence of one or more preferred frequencies. In particular, there
is nothing special in the distribution of frequencies at around 650 Hz,
which corresponds to an upper to lower kHz
QPO  frequency ratio of 1.5.

\subsection{Long term evolution}

Our data set also provides a unique view of the long term evolution
of a persistent neutron star LMXBs. We found that the $\sim 30-40$
day periodicity seen in the RXTE/ASM light curve of 4U 1636--53 
up to December 2004 (Shih
et al. 2005) is still present in our observations, up to Sepember 2006. 
Figs.\ \ref{licu} and \ref{hid} show
that 4U 1636--53 moves regularly along the HID track, following this
$\sim$30--40 days periodicity. In other words, the periodicity in the
light curve is caused by a regular alternation of soft (island) and
hard (banana) states. In combination with RXTE/ASM light curves (see
Shih et al. 2005), our observations suggest that the state
transitions have been repeating regularly for at least 5 years. 

The HID of 4U 1636--53 shares some similarities with the
corresponding diagrams for neutron star and black-hole transients
(Maccarone \& Coppi 2003; Belloni et al. 2005; Homan \& Belloni
2005). In particular, during outbursts, these systems follow  a counter-clockwise
path in their HID as they switch between their hard and soft states.
For these systems, the outbursts start with a
hard peak, which means that the hysteresis can also be interpreted as
the soft flux lagging the hard one (see e.g. Yu et al. 2004). The case
of 4U 1636--53 is the same. 
The hard-soft (island-banana) transition always takes place at a higher
flux value than the returning transition to the hard (island) state, similar
to what is observed in black-hole transients, where the hard-soft transition
always takes place at a higher flux than the reverse transition.
4U 1636--53 is the first persistent system for which such a
hysteresis cycle has been observed repeatedly. 
A comparison of the ASM light curve of 4U 1705--44 with the 
color analysis of Olive, Barret \& Gierli\'nski (2003) indicates that
this source might behave in a similar way.
Shih et al. (2005) show a similar quasi-periodic modulation in the
ASM flux of KS 1731--260, but no dense sampling with the PCA
is available to check for state transitions.
A more detailed
analysis of the hysteresis behavior will presented in a future paper.

\section{Conclusions}

We reported the results of a one and a half year campaign
to observe 4U 1635-53. The observational strategy of a regular monitoring
has the advantage of
yielding an unbiased coverage of the evolution of the system along
its quasi-regular 30-day oscillation. We found that this oscillation
corresponds to a regular alternation of island and banana states. We
traced the evolution of these oscillations in a hardness-intensity 
diagram and found that the source follows a hysteresis cycle similar
to that observed from transient black-hole and neutron-star LMXBs. 
In many observations, we detected kHz oscillations, most of which we
could classify as lower kHz QPO. 
Their frequency distribution is rather flat and it is significantly different
from the frequency distribution found by BMH05, based on previous RXTE
observations. The conclusion is that there is no preferred frequency 
or frequency ratio in 4U 1636--53.

\section*{Acknowledgments}

TB acknowledges financial contribution from contract ASI-INAF  I/023/05/0.

\label{lastpage}

\end{document}